\documentclass[12pt]{article}
\newcommand{\etal}{\textsl{et al.}}
\newcommand{\fcc}{$fcc\ $}
\newcommand{\hcp}{$hcp\ $}
\newcommand{\fcchcp}{\mbox{$fcc$-$hcp\ $}}
\newcommand{\ds}{\Delta s}
\newcommand{\dsstar}{\ds^*}

\newcommand{\conf}{\{ \vec{u} \}}
\newcommand{\sM}{\mathcal{M}}
\newcommand{\wgts}{\{ \eta \}}

\newcommand{\vSigma}{\vec{\Sigma}}

\begin{document}
\title{Stacking Entropy of Hard Sphere Crystals}
\author{Siun-Chuon Mau and David A. Huse \\
Department of Physics: Joseph Henry Laboratories \\
Jadwin Hall \\
Princeton University \\
Princeton NJ 08544 \\
U.S.A.}
\date{\today}
\maketitle

\begin{abstract}
Classical hard spheres crystallize at equilibrium at high enough density.
Crystals made up of stackings of 2-dimensional hexagonal close-packed layers
(e.g. \fcc, \hcp, etc.) differ in entropy by only about $10^{-3}k_B$ per
sphere (all configurations are degenerate in energy). To readily resolve
and study these small entropy differences, we have implemented
two different multicanonical Monte Carlo algorithms that allow direct
equilibration between crystals with different stacking sequences. 
Recent work had demonstrated that the fcc stacking has higher entropy
than the hcp stacking.  We have studied other stackings to demonstrate
that the fcc stacking does indeed have the highest entropy of {\it all}
possible stackings.  The entropic interactions we could detect involve
three, four and (although with less statistical certainty) five consecutive 
layers of spheres.  These interlayer entropic interactions fall off in
strength with increasing distance, as expected; this fall-off appears to
be much slower near the melting density than at the maximum 
(close-packing) density.
At maximum density the entropy difference
between \fcc and \hcp stackings is $0.00115 \pm 0.00004 \, k_B$ per sphere,
which is roughly 30$\%$ higher than the same quantity measured near the
melting transition. 
\end{abstract}


\section{Introduction}

	The question of which crystalline
stacking of hard spheres near close
packing is thermodynamically stable is a long standing
one. The interest is partly due to the extreme anharmonicity of hard
core interactions and partly due to the \fcchcp phase 
transition in solid helium~\cite{He}.  This problem is difficult both
experimentally and theoretically.  Experimentally, classical hard spheres are
approximated by spherical colloidal particles with 
interactions whose ranges are very short
compared to their radius.  Deviations from ideal hard spheres are due
to polydispersity of the spheres, and due to interactions.
The \mbox{van der Waals} interaction can be 
reduced by matching the dielectric coefficients
of the particles and the solvent~\cite{vander1,vander2}.
Since, for ideal hard spheres, the free energy differences between the
different stackings are very small, as we will see, one would expect
the equilibration time to be very long.  Most studies have
seen a random stacking pattern.
However, some experiments have reported that the observed random
stacking patterns in slowly-grown or well-annealed
colloidal crystals are biassed 
more towards \fcc rather than \hcp stacking~\cite{Pussey}.

	The free energy difference between different classical hard sphere
crystals at fixed volume is only due to the entropy difference, since the
energy is the same for all allowed configurations.  The numerical work,
before the present paper, had only looked at the \hcp and \fcc
stackings.  The first studies calculated pressure using molecular
dynamics simulations~\cite{moldyns} and then obtained entropies by
integrating the pressure {\it vs.} volume along reversible paths from states
with known entropy~\cite{eqnstate,wrong,singlet,dodecahedra}.  
These studies were not able to detect the entropy difference  
between \fcc and \hcp crystals.  Later, Frenkel and
Ladd~\cite{einstein} instead integrated along a path 
connecting the hard sphere model to
Einstein crystals of the same lattice structure, by adding 
to the model ideal springs tethering each ball to its lattice site. 
In this approach they integrate the derivative of 
free energy with respect to
the spring constant. They came up with the bounds on the 
entropy difference per sphere:
\mbox{$ -0.001 < \dsstar < 0.002$} in units of $k_B$, where 
\mbox{$\dsstar \equiv s_{fcc} - s_{hcp}$}. 

	Recently, Bolhuis, Frenkel and the present authors used both
a new implementation of the 
multicanonical Monte Carlo (MCMC) method~\cite{MCMC1,MCMC2}, as well
as the Einstein crystal method to reduce the statistical errors down
to the $10^{-4} k_B$ per sphere level.  This allowed us to
accurately resolve the entropy difference
of roughly $10^{-3} k_B$ per sphere between \fcc and \hcp crystals, with
the \fcc crystal having the higher entropy.~\cite{comment}  This 
quantitatively corrected the recent pressure-integration study of
Woodcock~\cite{Woodcock}, confirming his result that the \fcc
crystal has higher entropy.  More recently, Bruce \etal \ have
found a superior implementation of the multicanonical
method for this problem, reducing the statistical error in $\dsstar$
down to near the $10^{-5} k_B$ level~\cite{Wilding}.  
The various results for $\dsstar$ are summarized in
Table~\ref{table:positive}.  It
is clear from the table that MCMC is able to obtain substantially
smaller statistical errors for this problem, 
compared to the more conventional integration
methods. 

	The reason why integration methods were used 
initially was that it 
did not appear possible for the hard-sphere system to equilibrate
between the \fcc and \hcp crystal structures, due to the very large,
or even infinite, free energy barrier separating them.  This was
certainly true for standard molecular dynamics or Monte Carlo methods.
However, the MCMC method, the implementations of which we will
summarize below, is designed precisely to eliminate such large
free energy barriers and allow equilibration between very different
states.  This permits a direct measurement of the relative
entropies of the two states simply by comparing the probabilities
of their occurrences in a single simulation.

	Since only the \hcp and \fcc crystals had been examined
in previous work, we have also looked at other stackings of
hexagonally close packed planes of spheres to make certain
that neglecting the other possible stackings was
reasonable.  The entropy differences
between the stackings can be described as due to interactions between
layers.  We have been able to detect the 
entropic interaction between a given
layer and its nearby layers that are two, three,
and possibly even four layers away.
These interactions are all of the sign that favors the \fcc stacking,
so we confirm, as is no surprise, that the \fcc 
stacking has higher entropy than {\it all}
other stackings, not just higher than \hcp.  
At the maximum packing density, we find that the interaction with
the third-neighbor layer is roughly an order of magnitude smaller
than that with the second-neighbor layer, as seems quite reasonable.
For lower density, near the melting transition, the fall-off of
these entropic interactions with distance appears to be much
slower, presumably reflecting the larger fluctuations of the
individual sphere positions. 

	For any given stacking, the entropy varies as a 
function of any homogeneous lattice
deformation at \emph{constant} volume fraction.  For the \fcc stacking,
the undeformed lattice has cubic symmetry, so must by symmetry be
a stationary point of the entropy vs. deformation, and it is the maximum.
For the other stackings, there is no such symmetry, and the maximum
entropy may be obtained for a deformation where the 
expansion of the lattice away
from close-packing is not isotropic.  We have looked for this possible
effect in the \hcp stacking by measuring the entropy vs. the uniaxial lattice
deformation (the $c/a$ ratio).  If there is an anisotropy, we were unable to
detect it.  If there is an entropy difference between the highest
entropy state and the isotropically-expanded state for the \hcp stacking,
this difference is less than $10^{-5}k_B$ per sphere, so is well below the
statistical errors in our comparisons to the other stackings.

	Another issue that arises in simulating hard-sphere
crystals is whether collisions between spheres that are not 
nearest-neighbors can be neglected.  It is certainly more convenient
to make the approximation of including the hard-sphere interaction
only between nearest-neighbors.  Of course this approximation is
terrible in the liquid phase, but we have found that it is actually quite
good in the solid phase even at the melting density.  There
we detected no difference
in  $\dsstar$ between the model where only nearest-neighbor spheres interact
and the model where second-neighbors also interact, indicating that
the difference is also smaller than our statistical errors
when comparing the entropies of different stackings.
It is usually not clear in the literature whether or not 
further neighbor interactions have been taken into account. 
This result of ours shows that it doesn't matter within the
solid phase at the presently available resolution of
the entropy.

\section{Model}
  \label{section:model}
  
  	The model we study is hard spheres: classical monodisperse
spheres that are forbidden to overlap.  All permitted configurations
(with no overlaps between spheres) have the same energy, which we may
set as zero energy.   At high enough density this system crystallizes at
equilibrium, and it is this crystalline equilibrium phase that we study here.
Some of our results are for the maximum possible, 
or close-packed density, which means the
system is being treated perturbatively to lowest order in the difference between the
density and the close-packed density.  In this limit the system is equivalent
to a simpler system of aligned, hard dodecahedra~\cite{dodecahedra}.

	We consider close-packed crystal structures that consist
of planes of hexagonally
close-packed (in 2-dimensions) spheres stacked up in the vertical
direction.  As is standard in discussing close-packed crystals, the stacking
sequence can be denoted by a sequence of the letters A, B and C, with
nearest-neighbor layers in the sequence always having different letters.
Any global permutation on the letters in the sequence simply represents
a rotation, reflection or translation of the structure, so will not change the entropy.
ABCABCA... is a sequence that represents the \fcc stacking, while ABABABA...
represents the \hcp stacking.  To fully remove the degeneracy associated with
the permutations, we may assign a Ising-like spin $\sigma_i$ to each
layer $i$ based on the local stacking sequence of that layer and its
nearest-neighbor layers immediately above and below it.  If that local  
stacking matches the \fcc pattern (e.g., ABC), $\sigma_i = +1$, while
$\sigma_i = -1$ if it instead matches 
the \hcp pattern (e.g., ABA).   For the three
layers being compared, all local stacking
patterns are equivalent to one of these two under permutations
of the labels A, B and C. This maps each stacking sequence on to a spin pattern of a
one-dimensional Ising model. 
  
	The entropy of a given stacking is a function of the
stacking sequence, which is described by the spins $\sigma_i$.
It is reasonable to expect that the shortest-range entropic
interactions are the largest, so the total entropy may be expanded as
\begin{equation}
	\label{eqn:model}
	S \ = \ Ans_0 \ + \ Ah \ \sum_{i=1}^n \ \sigma_i \ + \ 
                AJ \ \sum_{i=1}^n \ \sigma_i \sigma_{i+1} \ + \ 
                AJ' \ \sum_{i=1}^n \ \sigma_i \sigma_{i+2} \ + \
                Ah' \ \sum_{i=1}^n \ \sigma_i \sigma_{i+1} \sigma_{i+2} \ + \ \cdots ,
\end{equation}
for a stack of $n$ layers containing $A$ spheres per layer.  We have
periodic boundary conditions so $\sigma_{n+1} = \sigma_1$, etc.  The bulk of
the entropy is independent of the stacking sequence and given by $s_0$ per sphere;
$s_0$ is strongly density-dependent.  The shortest-range entropic
interaction is $h$, which involves the sequence over three consecutive
layers of spheres; this is the shortest sequence that can have distinct stackings.
This term is the magnetic field in the corresponding one-dimensional
Ising model.  The next term is the interaction $J$ between adjacent
spins, and arises from the entropic interactions among four consecutive
layers of spheres that are not already captured by the first term $h$.  We find, as is
reasonable, that $J < h$.  The next longer-range 
interactions ($h'$ and $J'$) that involve
five consecutive layers are also displayed above;
only in one case ($h'$ for density near melting) could we detect these interactions 
in our simulations at a level that may be statistically significant.

	The intuition behind this model is that the entropy of a sphere
is mostly determined by how it is caged by its nearest neighbors and to a
progressively lesser extent by the further neighbors.   The interaction parameters
do depend on the density.  We find that, in units of 
$10^{-5} k_B$ per sphere, the entropic interactions that we could detect
change from $(h,J) \cong (55,6)$ at the highest density (close-packing) to 
$(h,J,h') \cong (37, 18, 9)$
at the lowest density that the equilibrium crystal can have before it melts (at roughly
74\% of the close-packed density).   All detected interactions are 
of the sign such that the
\fcc stacking has the largest entropy.

\section{Multicanonical Monte Carlo Method}
  \label{section:methods}
  
  	To make direct comparisons of the entropies of hard-sphere
crystals with different stacking sequences, we want an algorithm that will
produce direct equilibration between the two sequences.  Then the
entropy difference is simply the logarithm of the ratio of the equilibrium
probabilities of the system exhibiting the two sequences in question.
Of course, near close-packing the hard spheres in a physically realistic
molecular dynamics or Monte Carlo simulation are strongly trapped by their
neighbors, so the stacking pattern will not change in any reasonable
time scale.  The multicanonical method ~\cite{MCMC1,MCMC2} was invented
to allow systems to transform at equilibrium between states that are separated
by a high free energy barrier.  This method has been generalized and applied to this
hard-sphere crystal problem in two ways, which we describe next.

	In both implementations the position of sphere $i$ is
given as ${\bf r_i  =  R_i  +  u_i}$, where ${\bf R_i}$ is the ideal reference lattice
position in the absence of fluctuations and ${\bf u_i}$ is the displacement
of sphere $i$ away from that reference position.  The algorithms both have Monte Carlo
moves that change the reference lattice without changing the displacements
${\bf u_i}$, as well as more standard moves that move individual spheres
without changing the reference lattice.
We describe the shear implementation first; this method we developed and used
to obtain our first results ~\cite{comment}.  However, 
Bruce, \etal ~\cite{Wilding} subsequently developed the simpler
overlap implementation, which we find
is computationally more efficient and easier to program, so we used it
for all of our more recent simulations.

\subsection{Shear Implementation}
  \label{subsection:shear}

	In the shear implementation we used an
equally-spaced sequence of ideal
reference lattices, ${\bf R_i}(\lambda)$, labelled by an index, 
$\lambda = 0,1,2,...,h$, that linearly interpolate
between the two stackings of interest, which are the beginning ($\lambda = 0$)
and the end ($\lambda = h$) of that sequence.  Thus we have 
\begin{equation}
	h{\bf R_i}(\lambda) = (h-\lambda){\bf R_i}(0) + \lambda{\bf R_i}(h) .
\end{equation}
This produces a $\lambda$-dependent relative shear between each pair
of adjacent layers whose local stacking pattern changes.
The reference sites in the intermediate lattices ($0 < \lambda < h$) are not
all equally spaced, and generally some are too close together for the hard
spheres to fit without touching.  In the original model, two spheres are assumed to touch if
their separation, $|{\bf r_i - r_j}|$, is less than $d$, the diameter of a sphere,
and this must remain true for the two stackings of interest, which are
represented by $\lambda = 0$ and $h$.
For the intermediate lattices, on the other hand, we allow the distance of contact, 
$d_{ij}(\lambda)$ to be either larger or smaller for
each pair of nearby spheres in adjacent layers 
whose relative reference position, ${\bf R_i - R_j}$,
changes with changing $\lambda$.  We attempt, using feedback, 
to choose these $d_{ij}(\lambda)$ so that
the entropy is a monotonic function of $\lambda$ and the average displacements,
$<{\bf u_i}>$ vanish for all $i$ and $\lambda$.  Note that the pairwise interactions
are different for each interpolation point, $\lambda$.  This is different from the
original MCMC method ~\cite{MCMC1,MCMC2}, where the Gibbs distribution
is multiplied by a $\lambda$-dependent
(but otherwise configuration-independent) reweighting factor, in order to
make the free energy monotonic between the two states of interest, thus
eliminating the free energy barrier.
	
	To start simulating one has to choose how many interpolation
points to use, $h$, and values for the $d_{ij}(\lambda)$.  There are two types of moves.
One is a single sphere move, changing one of the displacements ${\bf u_i}$.  
The other is
a $\lambda$-move that increases or decreases $\lambda$ by one without
changing any of the sphere displacements.  Any attempted move of either type 
is accepted if it does not result in any contact between spheres.  The entropies
and average displacements, as well as the acceptance rates of the moves are measured.
Based on these results $h$ is adjusted to attempt to 
maximize the rate of equilibration between the two stackings of interest, and the
$d_{ij}(\lambda)$ are adjusted to attempt to make 
the entropy monotonic and eliminate the
average displacements.  If this feedback is successful, we then measure the
relative entropy of the two stackings.

	We succeeded in getting this procedure to work for comparing the
\fcc and \hcp stackings for lattices of size up to $8^3$.  However, the difficulty of
getting the feedback to converge appeared to be increasing strongly with lattice
size.  Typically, ``bottlenecks'' would form 
between the two stackings of interest
where the $\lambda$-move acceptance rate was very small or zero, preventing
equilibration, and the attempts at eliminating these bottlenecks through the
feedback were time-consuming and not always succesful.  However, we were
able to obtain the entropy difference between the \fcc and \hcp stackings to within
statistical errors of roughly $10^{-4} k_B$ per sphere ~\cite{comment},
as is summarized in Table 1.  Then we
learned of the much more staightforward overlap implementation of MCMC for
this problem reported by Bruce, {\it et al.} ~\cite{Wilding}, which we discuss next.

\subsection{Overlap Implementation}
  \label{subsection:overlap}
  
  	The overlap implementation of MCMC ~\cite{Wilding} uses only
two reference lattices, which are the two 
different stackings whose entropy we
are comparing.  Let us call these two reference lattices $\alpha$ and 
$\beta$.  Again there are standard single-sphere moves and
changes of the reference lattice.  For any reasonable sized lattice,
the move that changes reference lattices will be rejected due to
sphere overlaps for all but an infinitesimal fraction of the
sphere configurations.  What is needed is to bias the simulation towards
those rare sphere configurations that allow the stacking to be changed.

	To do this, Bruce \etal \ ~\cite{Wilding}
introduce the overlap order parameter
\begin{eqnarray} 
	\sM(\conf) \equiv M(\conf,\alpha) - M(\conf,\beta)
\end{eqnarray}
where $M(\conf,\gamma)$ is the number of pairs of spheres that overlap in the
configuration 
$\conf$ for stacking $\gamma$.  For any allowed configuration
with stacking $\gamma$, $M(\conf, \gamma) = 0$, but for configurations
of the other stacking, usually $M(\conf, \gamma) > 0$.  To have the
change of stacking be an allowed move, we need $\sM(\conf)=0$, so no
overlaps are produced by the move that changes the reference lattice.

	The overlap multicanonical simulation samples the biassed, 
but unnormalized, distribution 
\begin{eqnarray}
	P(\conf, \gamma \mid  \wgts) \equiv
	P(\conf, \gamma ) \ e^{\eta(\sM(\conf))},	
\end{eqnarray}
where $P(\conf, \gamma)$ for (unbiassed) hard spheres is simply a constant for
all allowed sphere configurations in stacking $\gamma$, and is
zero otherwise.  The weights
$\wgts$ are chosen to eliminate the free energy barrier separating
the two stackings, thus allowing equilibration between them.

	Let $P(\sM)$ be the equilibrium, normalized, unbiassed
($\eta(\sM)=0$ for all $\sM$) probability distribution
of the overlap, assuming the system does fully equilibrate between
the two stackings.  Then the probability of being in stacking
$\alpha$ is
\begin{eqnarray}
	P_{\alpha} = \frac{P(0)}{2} + \sum_{\sM<0} P(\sM),
\end{eqnarray}
and the entropy difference we are interested in is
\begin{eqnarray}
	S_{\alpha} - S_{\beta} = k_B \ln \left( \
	\frac{P_{\alpha}}{1-P_{\alpha}} \ \right)  .
\end{eqnarray}
$P(\sM)$ has a local minimum at $\sM = 0$ and local maxima
at positive and negative $\sM$ that represent the most
probable overlaps for the two different stackings.  The weights 
$\eta(\sM)$ are chosen to be nonzero only between these two
local maxima of $P(\sM)$.  The unnormalized, biassed probability
distribution for $\sM$ is
\begin{eqnarray}
	P(\sM \mid \wgts) = P(\sM) \ e^{\eta(\sM)} .
\end{eqnarray}
We choose the weights $\wgts$ so that $P(\sM \mid \wgts)$
is linear in $\sM$ between the maxima of $P(\sM)$.  A simulation
with a given set of weights produces estimates of
$P(\sM)$ and also a new estimate of what the appropriate
weights are to achieve this linearity.  
These new weights are then used for the next,
longer simulation if the statistical errors have not yet been
reduced down to the desired level.  This procedure 
straightforwardly and effectively
eliminates the free energy barrier between the two stackings and
allows an accurate measurement of the entropy difference.

The overlap implementation of MCMC does not suffer from 
the tendency to form ``bottlenecks'' that slowed down the
equilibration between the two stackings in our shear 
implementation of MCMC. 
We used the overlap method to obtain most of the data reported in this paper.
Where we compared the two implementations 
the measured entropy differences were, of course, the same.

\subsection{Boundary Conditions}
  \label{subsection:BC}
	
	Suppose one stacks $N_3$ planes of 2-dimensionally 
(hexagonally) close-packed
spheres to form an arbitrary stacking. Each plane has $N_1 \times N_2$
spheres~\footnote{We always choose $N_1 = N_2$ to preserve the
hexagonal in-plane symmetry}, $N_1$ in the $i$-direction and $N_2$ in the
$j$-direction. The $i$-direction is chosen to coincide the the $x$-direction
and the $j$-direction is chosen at $60^\circ$ anti-clockwise from the
$x$-direction looking from above. $i$ and $j$ are the two basis directions 
of the 2-dimensional close packing. Each site has coordinate $(i, j)$ 
in-plane, $i = 0, 1, \ldots, N_1 -1 \ \mathrm{and} \ j = 0, 1, \ldots, N_2
- 1$. 

	Any stacking can be formed by fixing the position of the $(i,
j,k)$ reference site (in layer $k$) relative to the \textsl{same} 
site $(i,j,k\pm1)$ in the nearest layers. Define
$\vSigma_k = \vec R (i,j,k+1) - \vec R (i,j,k)$, where $\vec R$ are the
reference sites.  We take our unit length to be the lattice spacing
so  $|\vSigma_k| = 1$.
We use
\begin{eqnarray}
  \label{ourBC}
  \vSigma_k = \left\{ 
    \begin{array}{ll}
      \vSigma_+ \equiv (\frac{1}{2}, \frac{1}{\sqrt{12}},
	\sqrt{\frac{2}{3}}) & \mbox{\small{if from layer $k$ to layer $k+1$ is a}} \\
  	   & \hspace{4 mm} \mbox{\small{forward permutation of $ABC$}}  , \\
      \vSigma_- \equiv (\frac{1}{2}, - \frac{1}{\sqrt{12}},
	\sqrt{\frac{2}{3}}) & \mbox{\small{if from layer $k$ to layer $k+1$ is a}} \\
  	    & \hspace{4 mm} \mbox{\small{backward permutation of $ABC$}}  . \\
    \end{array}
  \right.
\end{eqnarray}
Taking $\vSigma_k = \vSigma_+$ for all $k$ gives an \fcc stacking.
For the \hcp stacking $\vSigma_k = \vSigma_+$ for $k$ even and
$\vSigma_k = \vSigma_-$ for $k$ odd (or {\it vice versa}).


	All simulations are done with periodic BC which is implemented
in the usual way: 
\begin{eqnarray*}
	\mbox{sphere at site } (i, j, k) & = & 
	\mbox{sphere at site } (i + N_1, j, k) \\
	& = & \mbox{sphere at site } (i, j + N_2, k) \\
	& = & \mbox{sphere at site } (i, j, k + N_3)  , \\
\end{eqnarray*}
so $\vSigma_k = \vSigma_{k+N_3}$. Our implementation of 
periodic BC allows any $N_3$ which is a
multiple of both of the periods of the $\vSigma_k$ patterns of the two
stackings in a given simulation.  For example, the period for \fcc is
one layer, while that for \hcp is two layers, so $N_3$ can be any
even number when we compare \fcc and \hcp entropies.

\section{Results}
  \label{section:results}

	Our results for entropy differences between different stackings at
both close-packing ($\rho/\rho_{cp} = 1$) and 
near melting ($\rho/\rho_{cp} = 0.739$) for different system sizes are 
summarized in Table~\ref{table:stackings}. 
A statistically significant finite-size effect was detected only
for the smallest size $4^3$.  Based on this, we assume that the finite-size
effect is negligible for sizes $8^3$ and larger, and we use those data for
calculating the values of the entropic interactions using our
Eq.\ref{eqn:model}.

	At close packing we fit, using the results of $N = 8^3, 9^3, 
\ \mathrm{and} \
10^3$ and the various different stackings, 
first letting all four parameters $\{h, J, J', h'\}$ 
vary, and then setting $h'=J'=0$ and only varying
$\{h, J\}$.  In the former case we obtain $h = 54.6 \pm 2.8, J = 6.1 \pm 1.6, J' =
-0.3 \pm 1.1 \ \mathrm{and} \ h' = 3.4 \pm 2.7$ in units of $10^{-5}
k_B$ with $\chi^2 / (df = 3) = 1.5$ and for the
$h'=J'=0$ fits: $h = 57.2 \pm 2.1 \ \mathrm{and}
\  J = 6.0 \pm 2.2$ with $\chi^2 / (df = 5) = 1.9$. 
We can see that the first fit with $J' \ \mathrm{and} \
h'$ allowed to vary gives values of them consistent with zero. 
Comparing the two fits we
also see that the inclusion of these two longer-range interactions
in the fit does not 
significantly perturb the
values of the shorter-range interactions $h$ and $J$. We therefore conclude that our
data can be explained using the model  
with \textsl{only} $h \ \mathrm{and} \ J$ non-zero. Indeed, $J' \
\mathrm{and} \ h' $ not being important is consistent with the small 
system size $N = 8^3$
being close to the thermodynamic limit. The signs of the non-zero 
interactions all favor \fcc stacking, therefore \fcc has higher entropy 
than all other stackings, consistent with experiment\cite{Pussey}. 
Notice that $h >> J >> J',h'$ shows that the entropic interactions decrease rapidly as 
their range increases.

	Similar fits of our data at a density ($\rho/\rho_{cp} = 0.739$) 
near melting yield $h = 36.9 \pm 3.1$, $J = 18.2 \pm 3.0$,
$J' = 2.5 \pm 2.2$ and $h' = 8.8 \pm 2.8$ with $\chi^2 / (df = 2) = 0.2$.
Thus it appears that the entropic interactions decrease in relative
magnitude with distance much more slowly at this lower density than they do near 
close-packing, which is perhaps expected, given the larger
free volume which allows the spheres to make larger
excursions away from their ideal lattice positions.  
Again, the interactions are all of the sign that
favor the \fcc stacking.  Our detection of $h'$ is only at the three
standard deviation level, so has a small chance of being just a
statistical fluctuation in the data.  However, if we fit assuming
$h'=J'=0$ the quality of the fit declines strongly, giving
$h = 43.2 \pm 3.7$ and $J = 17.3 \pm 3.8$ with 
$\chi^2 / (df = 4) = 5.4$.  


	For a general stacking pattern, the expansion as the density
is reduced from close-packing need not be isotropic.  For the \fcc
stacking it must, by the cubic symmetry, but for the other stackings,
the expansion along the direction normal to the layers can be
different from that along the directions parallel to the layers.
We have tested for this by allowing the ratio of these two expansions to vary
in a simple simulation of the \hcp stacking at
close-packing, measuring the entropy {\it vs.} the ratio,
and fitting to find the ratio that maximizes the entropy.  We find that this optimal
ratio is within $\pm0.002$ of isotropic, and the entropy difference
between isotropic expansion and the optimal expansion ratio is no
more than $10^{-5}k_B$ per sphere, so is smaller than the statistical
uncertainties in our simulations.  Because of this we have always
assumed isotropic expansion in the the entropy comparisons we have made.

	The issue of further neighbor interactions arises for
$\rho/\rho_{cp} < 1$.  At close-packing it suffices to test only for
collisions between nearest-neighbor spheres because further neighbors
cannot touch.  Not testing for further-neighbor collisions speeds up
the computer program.  As the density is reduced can we keep this
approximation?  For a model that includes only nearest-neighbor
interactions between spheres, the crystal is actually only metastable:
once a sphere ``escapes'' from the cage of its nearest-neighbors it
wanders freely.  We find that for densities at or above the melting
density, the rate of these ``escapes'' is very low,
allowing a good measurement of the entropy of the 
(now metastable) crystal.
We have also measured the entropy differences between the model
with only nearest-neighbor interactions and the otherwise identical
model with nearest and next-nearest-neighbor interactions, $S_n -
S_{nn}$, near melting (see Table~\ref{table:furtherneighbors}). 
Of course, adding the extra interactions does reduce the entropy a little
(roughly $8 \times 10^{-5} k_B$ per sphere), but this reduction is the
same, within errors, for both \fcc and \hcp stackings.  Thus we conclude
that any systematic error in our entropy comparisons due to using only
nearest-neighbor interactions are smaller than the statistical errors.
Therefore, we have used the faster nearest-neighbor only model in
most of our simulations near the melting density.

	

\newpage

\begin{table}
\begin{tabular}{|rcrcrcrcccl|} \hline \hline
$\rho / \rho_{cp}$  &&        $N$    &&   
\multicolumn{3}{c}{$10^{5} \times \Delta s^* / k_B$}  && method &&
ref. \\
\hline
0.736	&&  12000  &&  230  &&  (100)  && pressure integration && \cite{Woodcock} \\
0.736  &&  12096	   &&  87   &&  (20)  && Einstein crystal     &&
\cite{comment} \\
0.7778  &&  5832   &&   86  &&   (3)   && MCMC, overlap implementation
&& \cite{Wilding} \\
1.00	&&   1000  &&  113   &&   (4)  && MCMC, overlap implementation &&
present \\ 
1.00	&&   512    &&  110   &&   (20)  && MCMC, shear implementation  &&
present \\
0.731	&&   512    &&  85   &&   (10)  && MCMC, shear implementation &&
present \\
\hline \hline
\end{tabular}
\caption{Recent results of \fcchcp simulations for various densities (scaled
by the close-packed density $\rho_{cp}$).   $N$ is the number of spheres
in the samples simulated.  The entropy difference per sphere is $\Delta s^*$,
with the statistical errors in parenthesis.  (\fcc has higher entropy.) 
Please note that the errors are particularly small for the
overlap implementation of MCMC developed by Bruce, {\it et al.} }
\label{table:positive}
\end{table}

\begin{table}
\begin{tabular}{|rrcccccc|} \hline \hline
$ \rho/\rho_{cp}$ & $N$    &  \multicolumn{3}{c}{$\Delta s$}   & 
		\multicolumn{2}{c}{$10^{5}\times \Delta s/k_B$}       &\\ 
\hline
1     & $4^3$  & $s_+ - s_-$                    & = & $2 h  +  2 h'$ 
							& 91     & (5) &\\
1     & $6^3$  & $s_+ - s_-$     		& = & $2 h  +  2 h'$          
							& 107  & (4) &\\
1     & $8^3$  & $s_+ - s_-$  	                & = & $2 h  +  2 h'$ 
							& 119  & (3) &\\
1     & $10^3$ & $s_+ - s_-$	                & = & $2 h  +  2 h'$         
							& 113  & (4) &\\
1     & $8^3$  & $s_{++--} - s_{+-}$		& = & $J  -  2 J'$
							& 6.1  & (1.5) &\\
1     & $9^3$  & $s_+ - s_{+--}$		& = & 
	$\frac{4}{3} h + \frac{4}{3} J  +  \frac{4}{3} J'$ 
						        & 82.6 & (2.7) &\\
1     & $8^3$  & $s_+ - s_{++--} $		& = & $h + J +  2 J'  + h'$   
							& 61.2 &  
(2.2)&\\
1     & $8^3$  & $s_{+-} - s_- $		& = & $h - 2 J +  h'$ 
							& 44   & (4) &\\
1     & $8^3$  & $s_{++++++--} - s_{++++-++-}$  & = & $\frac{1}{2} J + \frac{1}{8} J' + h'$
							& 8.2 & (2.2) &\\
0.739 & $8^3$  & $s_+ - s_-$			& = & $2 h + 2 h'$ 
							& 90.2 & (4.3) 
&\\
0.739 & $8^3$  & $s_{++--} - s_{+-}$            & = & $J - 2 J'$ 
							& 13.7 & (2.9) 
&\\
0.739 & $8^3$  & $s_{+-} - s_- $		& = & $h - 2 J +  h'$ 
							& 10.5 & (5.0) 
&\\
0.739 & $8^3$  & $s_{++++++--} - s_{++++-++-}$  & = & $\frac{1}{2} J + \frac{1}{8} J' + h'$
							& 19 & (3) &\\
0.739 & $8^3$  & $s_{+++++-+-} - s_{+++-}$      & = & 
		\vspace{0.5 mm} $\frac{1}{2} J' + \frac{1}{2} h'$
							& 6.2 & (3.2) &\\
0.739 & $8^3$  & $s_{++++-++-} - s_{++---++-}$  & = & $\frac{1}{2} h + \frac{1}{2} J' - \frac{1}{4} h'$
							& 18.0 & (1.9) 
&\\
\hline\hline
\end{tabular}
\caption{This table summarizes the entropy differences per sphere between various pairs of
stackings at the densities we studied.   
The subscript on $s$ is a sequence of $\sigma_i$'s that is
repeated to get the stacking sequence, so $+$ denotes
\fcc, $-$ denotes \hcp, and the others are less simple stackings (see text). }
\label{table:stackings}
\end{table}

\begin{table}
\begin{tabular}{|rrcrrc|} \hline\hline
$\rho/\rho_{cp}$   &  $N$    &   $\ds$     &    
		\multicolumn{2}{c}{$10^{5} \times \ds/k_B$}& \\
\hline
$0.739$   & $8^3$ & $s^{fcc}_n - s^{fcc}_{nn}$  &   8.3 & (1.9) &\\
$0.739$   & $8^3$ & $s^{hcp}_n - s^{hcp}_{nn}$  &   7.8 & (1.9) &\\
\hline
\end{tabular}
\caption{The entropy differences per sphere
between hard sphere crystals with only nearest
neighbor interactions ($n$) and the otherwise identical system with both
nearest and next nearest neighbor interactions ($nn$).  The systems with
added interactions have lower entropy, but the change is independent of
the stacking pattern at the present statistical accuracy.}
\label{table:furtherneighbors}
\end{table}

\end{document}